\documentclass[conference]{IEEEtran}
\IEEEoverridecommandlockouts

\usepackage{cite}
\usepackage{amsmath,amssymb,amsfonts,amsthm}
\usepackage{algorithmic}
\usepackage{graphicx}
\usepackage{textcomp}
\usepackage{xcolor}
\usepackage{tikz}  
\usetikzlibrary{spy}
\usepackage{color, colortbl}
\usepackage{pgfplots}
\pgfplotsset{compat=newest}
\usepackage{multicol}
\usepackage{circuitikz}
\usepackage[nolist]{acronym}
\usepgfplotslibrary{groupplots}
\usetikzlibrary{shapes}
\usepackage{siunitx}  
\usepackage{graphicx}
\usepgfplotslibrary{patchplots}
\usepackage{amsmath,amssymb}
\usepgfplotslibrary{fillbetween}
\definecolor{mittelblau}{RGB}{0, 126, 198}
\definecolor{violettblau}{cmyk}{0.9, 0.6, 0, 0}
\definecolor{rot}{RGB}{238, 28 35}
\definecolor{apfelgruen}{RGB}{140, 198, 62}
\definecolor{gelb}{RGB}{1, 221, 0}
\definecolor{orange}{RGB}{244, 111, 33}
\definecolor{pink}{RGB}{237, 0, 140}
\definecolor{lila}{RGB}{128, 10, 145}
\definecolor{hellgrau}{RGB}{224, 224, 224}
\definecolor{mittelgrau}{RGB}{128, 128, 128}
\definecolor{dunkelgrau}{RGB}{80,80,80}
\definecolor{anthrazit}{RGB}{19, 31, 31}
\definecolor{darkgreen}{RGB}{0.125,0.5,0.169}

\usepackage{ifthen}
\usetikzlibrary{matrix,calc}

\usepackage{hyperref}
\hypersetup{
    colorlinks=true,
    linkcolor=blue,
    filecolor=magenta,      
    urlcolor=cyan,
    breaklinks=true,
}

\ifCLASSOPTIONcompsoc
    \usepackage[caption=false, font=normalsize, labelfont=sf, textfont=sf]{subfig}
\else
\usepackage[caption=false, font=footnotesize]{subfig}
\fi

\def\BibTeX{{\rm B\kern-.05em{\sc i\kern-.025em b}\kern-.08em
    T\kern-.1667em\lower.7ex\hbox{E}\kern-.125emX}}

\newcommand\blfootnote[1]{%
  \begingroup
  \renewcommand\thefootnote{}\footnote{#1}%
  \addtocounter{footnote}{-1}%
  \endgroup
}

\newcommand*{\narrowfill@}[5]{%
  $\m@th\thickmuskip0mu\medmuskip\thickmuskip\thinmuskip\thickmuskip
  \relax#5#1\mkern-7mu%
  \cleaders\hbox{$#5\mkern-2mu#2\mkern-2mu$}\hfill
  \mkern-5mu %
  #4%
  \mkern-5mu %
  \cleaders\hbox{$#5\mkern-2mu#2\mkern-2mu$}\hfill
  \mkern-7mu#3$%
}
\newcommand*{\nLeftrightarrowfill@}{%
  \narrowfill@\Leftarrow\Relbar\Rightarrow\neq
}
\newcommand*{\xLeftrightarrow}[2][]{%
  \ext@arrow 3399\Leftrightarrowfill@{#1}{#2}%
}
\newcommand*{\xnLeftrightarrow}[2][]{%
  \ext@arrow 3399\nLeftrightarrowfill@{#1}{#2}%
}
\renewcommand{\qedsymbol}{$\blacksquare$}

\usepackage{balance}

\begin{document}

\title{Self-Supervised Radio-Visual\\ Representation Learning for 6G Sensing}

\author{
    \IEEEauthorblockN{
        Mohammed Alloulah\IEEEauthorrefmark{1}\IEEEauthorrefmark{2}, 
        Akash Deep Singh\IEEEauthorrefmark{4}\IEEEauthorrefmark{2}\IEEEauthorrefmark{3}, 
        Maximilian Arnold\IEEEauthorrefmark{2}
    }
    \IEEEauthorblockA{
        \IEEEauthorrefmark{2}Bell Labs
        \hspace{0.25cm} 
        \IEEEauthorrefmark{3}UCLA
    }
}

\maketitle

\begin{abstract}
In future 6G cellular networks, a joint communication and sensing protocol will allow the network to perceive the environment, opening the door for many new applications atop a unified communication-perception infrastructure.
However, interpreting the sparse radio representation of sensing scenes is challenging, which hinders the potential of these emergent systems. 

We propose to combine radio and vision to \emph{automatically learn} a radio-only sensing model with minimal human intervention. 
We want to build a radio sensing model that can feed on millions of \emph{uncurated} data points. 
To this end, we leverage recent advances in self-supervised learning and formulate a new \emph{label-free} radio-visual co-learning scheme, whereby vision trains radio via cross-modal \emph{mutual information}.
We implement and evaluate our scheme according to the common linear classification benchmark, and report qualitative and quantitative performance metrics.
In our evaluation, the representation learnt by radio-visual self-supervision works well for a downstream sensing demonstrator, and outperforms its fully-supervised counterpart when less labelled data is used.
This indicates that self-supervised learning could be an important enabler for future \emph{scalable} radio sensing systems.
\end{abstract}


\section{Introduction}

Imagine\blfootnote{\IEEEauthorrefmark{1}Correspondence to \href{mailto:alloulah@outlook.com}{alloulah@outlook.com}}\blfootnote{\IEEEauthorrefmark{4}Work was conducted when the author was under Bell Labs employment.} a world where cellular basestations are not only enabling untethered connectivity everywhere, but are also actively sensing the environment.
Supporting joint communication and sensing (JCAS) in next generation networks constitutes a step change in cellular systems functionality, and may well be what sixth generation (6G) networks are all about~\cite{wild2021joint,bourdoux20206g,Fettweis2021}.
That is, ubiquitous cellular basestations will allow us to detect the presence of objects of interest, estimate their shape, location, and movement patterns, whilst relying solely on radio signals.
With these newfound sensing capabilities, we will be able to gather urban analytics at unprecedented scales, e.g., for monitoring busy intersections in cities in support of safety-critical autonomous driving.

Concretely, 6G networks will enable a host of outdoor sensing applications ranging from traffic monitoring and parking spot detection, around-the-corner vehicle detection, to drone and pedestrian identification and localisation~\cite{wild2021joint}.
Given the privacy preserving advantage of radio sensing, and coupled with the ubiquity of cellular infrastructure, it is not unreasonable to expect that many city-scale analytics services would turn to radio spectrum for the provision of sensing primitives.

The sheer amount of sensory data produced by 6G basestations provides both opportunities and challenges for building scalable and robust sensing applications.
On the one hand, ubiquitous basestations can easily collect data from diverse environmental conditions, e.g., from dense urban or sparser suburban radio channels.
On the other hand, managing such data \emph{diversity} and \emph{scale} for producing highly performant sensing models becomes hard.
Particularly, state-of-the-art deep learning (DL) sensing models have typically required manual annotation of empirical data. For example, ImageNet, the flagship 14 million images dataset for object recognition, took 22 human years to annotate~\cite{NYU-DLSP21}.
Clearly, the progress of radio sensing should not hinge on the availability of \emph{laboriously annotated} data. 
Alternatively, radio sensing models could be trained on synthetic RF data using ray-trace simulations~\cite{singh2018simulation, hsiao2017ray, da2020cell}. 
Since the environment is simulated, we can specify the sensing scenario (i.e., label) first and then generate the corresponding data. 
However, synthetic data would only provide a 1st-order approximation of the real-world that is unlikely to meet the quality, scale, and richness requirements of highly-perfomant deep radio sensing models. 
Current synthetic data generation techniques for RF work well for simple environments, but their complexity-bounded performance degrades as we move to more realistic scenes which mirror the real-world~\cite{degli2014ray, lecci2020simplified, fuschini2017analysis}.

\begin{figure}[t]
  \centering
  \includegraphics{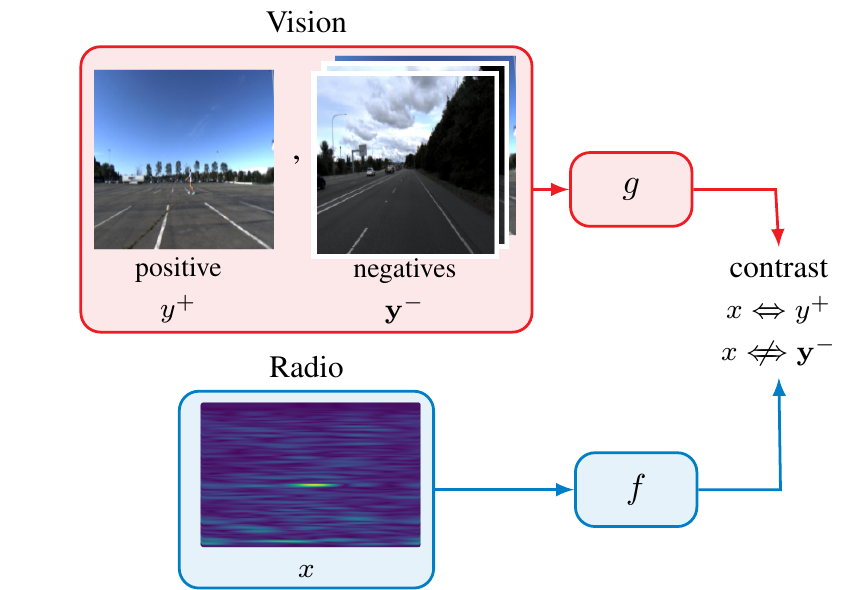}
  \vspace{-0.25cm}
  \caption{Tap into mutual information between paired camera and radar heatmaps to automatically learn a radio sensing model in a label-free fashion.
  This is achieved via contrastive radio-visual learning. Image $y^{+}$ is a positive pair of radar heatmap $x$ because they are recorded simultaneously, while other randomly-sampled images $\mathbf{y}^{-}$ are negative pairs.}
  \label{fig:radio-visual_learning}
\end{figure}

Recent advances in self-supervised learning have demonstrated a viable alternative to label-intensive supervised learning~\cite{bengio2013representation}.
Self-supervised methods provide two key advantages. Firstly, they do not require labelled data. Instead of solving a specific learning task, we solve an auxiliary task that helps the model learn (by \emph{proxy}) the underlying structure of the data. 
Secondly, self-supervised methods learn more granular and generalisable discriminative features than task-specific supervised methods~\cite{henaff2020data}.
This is because self-supervision generates higher information rate per sample than supervised learning as captured eloquently in Yann LeCun's cake analogy~\cite{lecun2019deep}.

In this paper, we draw inspirations from recent Natural Language Processing (NLP) breakthroughs~\cite{devlin2018bert,brown2020language} and Computer Vision (CV) successes~\cite{chen2020improved,henaff2020data} and propose a self-supervised method for radio representation learning. 
Specifically, we formulate an auxiliary prediction task that contrasts RF heatmaps to camera images. 
We show that such auxiliary \emph{contrastive} prediction promotes learning powerful radio neural networks that can be specialised for specific sensing applications. 
We argue that self-supervision for radio signals is an important enabler for emergent 6G sensing systems.

The paper describes the following contributions:
\begin{itemize}
    \item To the best of our knowledge, this paper is the first to demonstrate successful self-supervised representation learning for radio signals using cross-modal mutual information (MI) with camera images.
    A loss function bounded by MI enables such \emph{label-free} learning.
    \item We specialise the self-supervised radio model for a downstream task that classifies objects of interest in scenes from empirical measurements.
    \item We characterise our self-supervision-based model using a number of quantitative and qualitative metrics. When evaluated against a fully-supervised baseline, we find that the self-supervised model outperforms its supervised counterpart when less labelled data is used for training. 
\end{itemize}

\section{Overview \& Intuition}
Radio heatmaps are sparse representations of the environment. 
They are difficult to interpret and label by humans, because the underlying geometry of the physical space they capture is not straightforward to decipher.

Fig.~\ref{fig:radio-visual_learning} depicts our proposed radio-visual co-learning scheme.
The idea is to (i) match a radio heatmap with its corresponding groundtruth image and (ii) contrast this true radio-vision pairing against a number of false pairings using images sampled at random.
The very act of emphasising the \emph{contrast} between positive and negative radio-visual pairings gives rise to a robust learning signal as will be formalised and explained later.
Further, this simple yet powerful \emph{contrastive} learning can be self-administered, assuming only synchronised radio-visual measurements (i.e., true pairings). 
Hence, the \emph{contrastive} learning scheme is also \emph{self-supervised} and needs no labels.  

\section{Primer} \label{sec:primer}

We discuss next the foundational concepts of our self-supervised radio-visual learning.

\subsection{Self-supervised learning}

Unlike supervised learning where we need to manually annotate the data, self-supervised learning leverages intrinsic labels which can be ``contrived'' using the data itself. 
The most successful form of self-supervised learning to-date is in NLP, whereby a language model tries to predict randomly masked words in sentences.
Surprisingly, this simple contrived prediction acts as a signal that forces the network to learn fundamental aspects that characterise language altogether, e.g., word association and sentence-to-sentence context.
In addition to its groundbreaking applications in NLP~\cite{devlin2018bert,brown2020language}, self-supervised learning has also seen numerous successes in domains such as vision~\cite{doersch2015unsupervised} and multi-modal learning~\cite{arandjelovic2018objects}.
For example, visual self-supervised learning can be made to work by predicting whether randomly sampled image patches are related or not~\cite{doersch2015unsupervised}.  

\subsection{Contrastive loss} \label{sec:contrastive_loss}

\begin{figure}[t]
  \centering
  \includegraphics{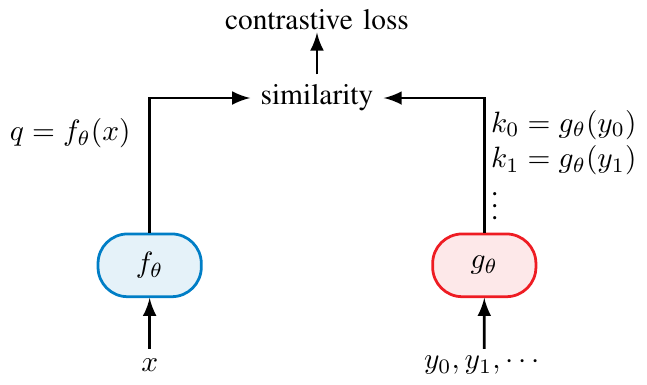}
  \caption{Contrastive loss for self-supervised learning.}
  \label{fig:contrastive_learning}
\end{figure}

For self-supervision to be effective, the signal that drives learning should also accentuate the difference (i.e. \emph{contrast}) between true and false predictions and their data samples.
The contrast between a positive datum and negative data is at the heart of learning good self-supervised models.
There are many incarnations to such a contrastive loss~\cite{oord2018representation,he2020momentum,afouras2020self}.
InfoNCE is one contrastive loss designed to drive learning in a way that preserves the mutual information (MI) between inputs~\cite{oord2018representation}.
Concretely, referring to Fig.~\ref{fig:contrastive_learning}, let $x$ and $y_i$ be two input signals encoded by two neural networks $f_{\theta}$ and $g_{\theta}$ such that $q = f_{\theta}(x)$ and $k_i = g_{\theta}(y_i)$, assuming some weights parametrisation $\theta$.
With each $x$, use $K+1$ samples of $y$ of which one sample $y^{+}$ is a true match to $x$ and $K$ samples $\{y_{i}^{-}\}_{i=0}^{K-1}$ are false matches. 
Then the constrastive loss $\mathcal{L}_c$ is~\cite{he2020momentum}
\begin{equation}
\mathcal{L}_c  = - \underset{x,y}{\mathbb{E}}\log \left[\frac{ \exp{( q \cdot k^{+} /\tau )} }{ \exp{( q \cdot k^{+} /\tau )} +\sum\limits_{ i } \exp{( q \cdot k_{i}^{-} /\tau )} }\right]
\label{eq:contrastive_loss}
\end{equation}
where $\cdot$ is the dot product operator, $k^{+/-} = g_{\theta}(y^{+/-})$ are encodings that correspond to true and false $y$ signals, vector $\mathbf{k}^{-} = \{k_{i}^{-}\}_{i=0}^{K-1}$ holds $K$ false encodings, and $\tau$ is a temperature hyper-parameter.

The contrastive loss of Eq.~\eqref{eq:contrastive_loss} has enabled a wave of progress in self-supervised image classification that inches ever closer and at times exceeds the performance of fully supervised methods.
Notably, SimCLR~\cite{chen2020simple} and MoCo~\cite{chen2020improved} are recent examples of bleeding-edge self-supervised vision systems.

The contrastive loss $\mathcal{L}_c$ has information-theoretic connections to the MI between $q$ and $k$ (vis-\`a-vis $x$ and $y$).
This can be seen noting the following.
First, the expression $\exp{( q \cdot k /\tau )}$ models a positive real score between $q$ and $k$, which is proportional to the (unnormalised) density ratio 
\begin{align}
\exp{( q \cdot k /\tau )} &\propto \frac{p(k|q)}{p(k)}
\label{eq:unnormalised_density_ratio}
\end{align}
Second, Eq.~\eqref{eq:contrastive_loss} has the form of a ($K+1$)-way softmax classifier that matches $q$ to its positive pair $k^{+}$.
Using Eq.~\eqref{eq:unnormalised_density_ratio} in Eq.~\eqref{eq:contrastive_loss}, van den Oord et al. show that $\mathcal{L}_c$ preserves the MI between $q$ and $k$~\cite{oord2018representation}.
Specifically, $\mathcal{L}_c$ becomes lower bounded by the MI between the encoded inputs\footnote{which are in turn lower bounded by the MI between the raw inputs} and that using more negative samples $K$ enhances the learnt representation~\cite{oord2018representation,poole2019variational}
\begin{align}
\mathcal{L}_c(q, k^{+}, \mathbf{k}^{-}) &\geq \log(K) - I(q;k) \label{eq_contrastive_loss_bound} \\
\mathcal{L}_c^{\text{optimal}}(q, k^{+}, \mathbf{k}^{-}) &= \log(K) - I(x;y)
\label{eq_contrastive_loss_opt}
\end{align}
where $I(q;k)$~\&~$I(x;y)$ denote respective MI's. For a self-contained treatment, we recapitulate the derivations of Eq.~\eqref{eq_contrastive_loss_bound}~\&~\eqref{eq_contrastive_loss_opt} in supplementary Sec.~\ref{sec:Lc_MI_connection}.

\subsection{Cross-modal co-supervision}

The notion of self-supervision can also be applied across different modalities for data with inherent multi-modal signals, e.g., video.
The sheer amount of video freely available on the internet makes this proposition increasingly appealing.
For instance, there is a body of literature that studies how representations for audio and vision can be learnt \emph{jointly}, e.g.,~\cite{arandjelovic2017look,afouras2020self} (and references therein).
Besides learning quality audio and vision features that are on par with supervised single-modal approaches, additional cross-modal tasks can be \emph{automatically} instantiated (from the joint embeddings), e.g., synchronisation and object localisation~\cite{afouras2020self,chen2021localizing}.

Similar to NLP and vision, audio-visual self-supervised systems use a contrived auxiliary task to promote cross-modal representation learning.
Earlier works dealt with predicting the \emph{correspondence} of an image and an audio spectrogram snippet using a binary cross-entropy loss, e.g.,~\cite{arandjelovic2017look}. 
Audio-visual correspondence in this context is static, i.e., from a single measurement snapshot.
More powerfully, recent works extend this correspondence to capture shared temporal dynamics between a video snippet and an audio snippet, e.g., ~\cite{afouras2020self}.

\section{Radio-Visual Contrastive Learning}

We next turn to discussing how to adapt the concepts treated in Sec.~\ref{sec:primer} for radio-visual self-supervised learning.

\subsection{Pre-training}

\begin{figure}[t]
  \centering
  \includegraphics{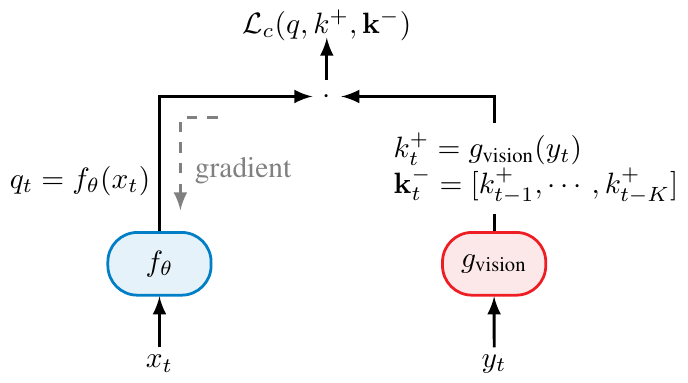}
  \caption{Radio-visual contrastive loss.}
  \label{fig:radio-visual_contrastive_learning}
\end{figure}

Our objective is to leverage the co-occurrence of events in the radio and vision domains and formulate a learning scheme that requires no labels.
That is, the onset of an event of interest across radio and vision will give rise to MI that can be captured and used as an automatic learning signal.
The judicious sampling of positive and negative data according to Eq.~\eqref{eq:contrastive_loss} extracts cross-modal MI.

\noindent \textbf{Data collation.} In practice, we operate a pair of camera and radar devices that are synchronised to continuously collect measurements.
The synchronisation timestamps allow us to construct the positive and negative data sample set $S_d = \{x_t, y_{t}^{+}, \mathbf{y}_{t}^{-}\}$ and their respective encodings $S_e = \{q_t, k_{t}^{+}, \mathbf{k}_{t}^{-}\}$, at a given time instance $t$.
The resultant dataset can be further filtered by \emph{light}\footnote{i.e., without laborious annotations such as bounding boxes} object detection algorithms in the vision domain to: concentrate on learning good encodings for specific objects of interest (e.g., pedestrians, cyclists, and cars), balance their distributions, etc.

\noindent \textbf{Architecture.} Our machine learning model consists of two parts: (i) a radar branch and (ii) a camera branch. During training, the two branches interact through the contrastive loss of Eq.~\eqref{eq:contrastive_loss}. 

\noindent \textbf{Representation learning.} Let ($x_t$, $y_t$) be a pair of a radio heatmap and a corresponding camera image, as depicted in Fig.~\ref{fig:radio-visual_contrastive_learning}.
A neural network with two branches for radio and vision ingests ($x_t$, $y_t$).
For the vision subnetwork, we use a pretrained convolutional model to encode $y_t$, yielding a positive encoding $k_{t}^{+} = g_\text{vision}(y_t)$.
The vision subnetwork is frozen and there is no gradient flowing back to it during training.
This aspect is different to self-supervised vision systems~\cite{chen2020simple,chen2020improved}.
The idea is to leave the vision encoder maximally consistent during training, which would significantly enhance the contrastive learning of the radar subnetwork.
Under our cross-modal settings, it also makes sense to derive learning from the higher-entropy vision modality.
We adopt a queue of negative encodings scheme proposed by He et al. in their momentum contrast (MoCo) system~\cite{he2020momentum}.
The negative encodings queue $\mathbf{k}_{t}^{-}$ is constructed by utilising past image encodings that do not correspond to the current radar heatmap\footnote{by construction of the dataset}, i.e., $\mathbf{k}_{t}^{-} = [k_{t-1}^{+}, \cdots, k_{t-K}^{+}]$.
The queue is a very efficient mechanism to enhance contrastive learning (cf. loss bound in Eq.~\eqref{eq_contrastive_loss_bound}) under practical compute and memory constraints.
Throughout training, a radar heatmap $x_t$ is encoded according to $q_t = f_{\theta}(x_t)$, and a gradient is computed to optimise for $q_t$'s similarity to its positive image pair $k_{t}^{+}$ and dissimilarity to $K$ negative images in $\mathbf{k}_{t}^{-}$.
This $K$-way contrastive learning can be trivially extended for the mini-batch settings of stochastic gradient decent (SGD).

\begin{figure}[t]
  \centering
  \includegraphics{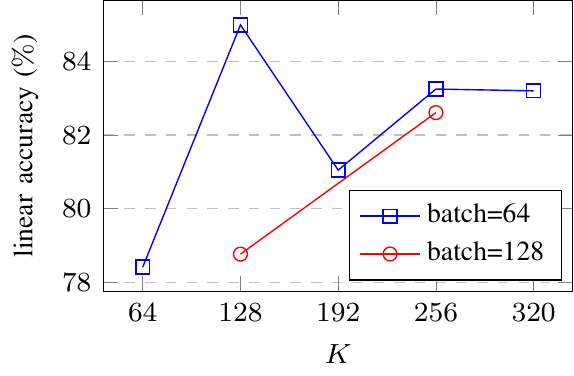}
  \vspace{-0.35cm}
  \caption{Linear classification as a function of number of negatives $K$, for two batch sizes. Generally, linear classification is enhanced with contrastive learning against more negatives, which agrees with the theoretical MI bound.}
  \label{fig:queue_size_effect}
\end{figure}

\subsection{Fine-tuning}

Contrastive representation learning concludes by arriving at a good $f_{\theta}$ than can used in a variety of sensing applications.
A task-specific model is then constructed using the output of $f_{\theta}$ as features combined with a classifier head.
Typically, sensing tasks also fine-tune (i.e., bias) these pre-trained representations towards application-specific criteria.

\subsection{Implementation details} \label{sec:implementation_details}

For both the vision and radio branches, we use pre-trained VGG-16 models~\cite{simonyan2014very}.
The VGG-16 model takes a image of $3 \times 224 \times 224$ dimensions as input.
For the radar branch, the heatmap is resampled and replicated $3$ times for a valid input.
The vision branch is frozen and only the radar branch is trained.
We slice the VGG-16 model at the 4th layer of its classifier stage to obtain a 4096-dimensional feature vector, i.e., $q, k \in \mathbb{R}^{4096 \times 1}$.
We use MoCo's shuffling trick to mitigate against batch normalisation (BN) statistical issues~\cite{he2020momentum}.
BN's statistical artefacts have also been observed in other works, e.g., ~\cite{labatie2021proxynormalizing,alloulah2021deep}.   

\section{Experiments}

\subsection{Dataset} 
We use the Camera-Radar of the University of Washington (CRUW) dataset~\cite{wang2021rodnet} for evaluating our radio-visual self-supervised learning.
The dataset contains around $47,000$ image-heatmap pairs in a variety of settings such as empty parking lots, driving on roads, and in front of buildings.  
Objects present in the dataset are: pedestrians, cyclists, and cars.
An example pair is depicted in Fig.~\ref{fig:radio-visual_learning}. 

\noindent \textbf{Downstream task.} We balance the class distribution of a subset of the dataset such that there are 4 non-overlapped classes in any given scene: empty, pedestrian, cyclist, and car.
The resultant 4-category sensing task maps onto the envisioned 6G use cases, e.g., pedestrian versus car discrimination for safety applications~\cite{wild2021joint}.
We use a standard 80/20 train/test split.

\begin{figure}[t]
  \centering
  \includegraphics{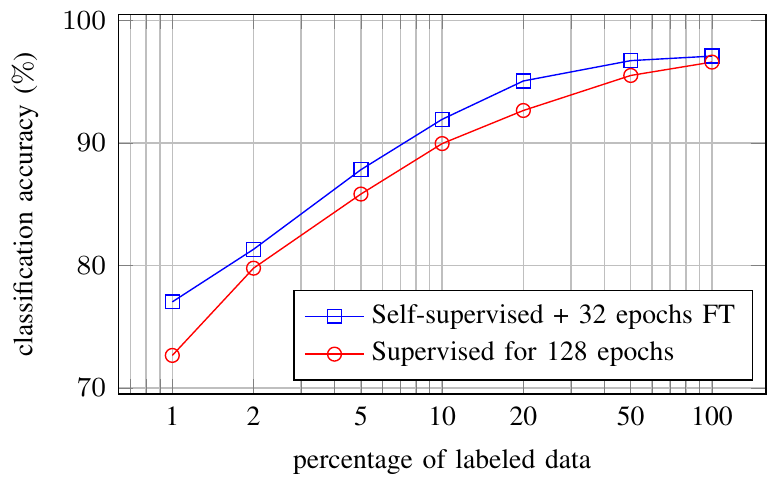}
  \vspace{-0.25cm}
  \caption{Efficiency measured w.r.t. accuracy versus percentage of labels used for fine-tuning (FT). 
  Self-supervised efficiency is compared to its fully-supervised counterpart.
  Self-supervised contrastive learning needs 32 epochs to \emph{exceed} the level of performance of fully supervised training using 128 epochs: a factor of $\mathbf{4\times}$ reduction in training.
  }
  \label{fig:efficiency}
\end{figure}

\subsection{Training}

Similar to MoCo~\cite{chen2020simple,chen2020improved}, we use an stochastic gradient descent (SGD) optimiser for contrastive pre-training.
The learning rate starts at 0.03 with an SGD weight decay of 1e-4 and momentum of 0.9.
The learning rate decays according to a cosine schedule.
We use two mini-batch sizes: 64 and 128.
We train in a distributed fashion on 4 GPUs for 400 epochs, which takes $\sim$12 hours for VGG-16.
For training the fully-supervised baseline, fine-tuning, and the linear classifier, we use a 1e-4 learning rate.
For the fully-supervised baseline, we train for 300 epochs.
For fine-tuning and the linear classifier, we train 32 epochs only.

\subsection{Evaluation protocol}

\noindent \textbf{Linear classification.} We benchmark the quality of our self-supervised radio representation learning using the common \emph{linear} classification protocol.
Under the linear classification protocol, we: (1) pre-train using contrastive learning, (2) freeze the learnt features, and then (3) train a supervised classifier with a \emph{linear} fully-connected (FC) layer whose output is normalised using a softmax.

\noindent \textbf{Efficiency.} We investigate the data and training efficiency of the contrastive representation $f_{\theta}$.
Data efficiency refers to the amount of labels required to build a downstream task using $f_{\theta}$.
Training efficiency refers to the number of epochs required to converge to a downstream representation with competitive performance.
Both metrics are evaluated against their fully-supervised counterparts.

\section{Results}

\subsection{Quantitative}

We begin by examining the linear classification performance of self-supervised radio features.

\noindent \textbf{Linear classification.} Eq.~\eqref{eq_contrastive_loss_bound} tells us that the contrastive loss benefits from increasing the number of negatives. 
We investigate the effect of queue size $K$ on the linear classification accuracy in Fig.~\ref{fig:queue_size_effect}.
For mini-batch size 64, we sweep $K$ from 64 to 320 in 64 increments.
For mini-batch size 128, $K \in [128, 256]$ only due to GPU resource constraints.
Generally, apart form a blip in mini-batch 64 at $K=128$, linear accuracy increases steadily with $K$ for both mini-batches.

\noindent \textbf{Efficiency.} Fig.~\ref{fig:efficiency} depicts the label efficiency of self-supervised features and compares it to the fully-supervised efficiency.
Using all dataset labels, both achieve a comparable accuracy of 97\%.
However, self-supervised features beat the supervised ones when we decrease the number of labels used for training.
For instance, with 1\% labels, we see a gap of about 3.5\% in favour of self-supervision.
This observation is generally in agreement with recent findings from the CV domain~\cite{henaff2020data}, which report on even more drastic advantages to self-supervision in the ``low label density'' regime.

\begin{figure}[t]
\includegraphics{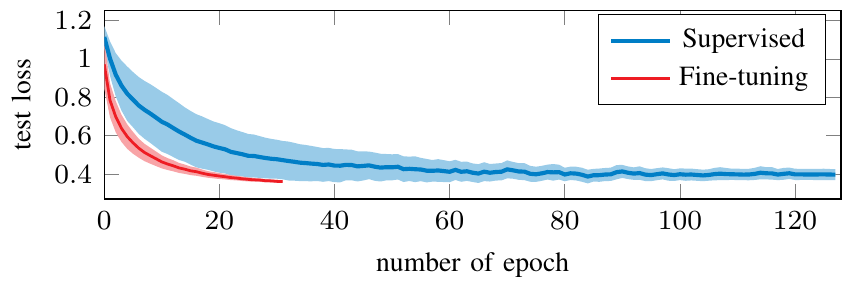}
\vspace{-0.35cm}
\caption{Comparing the testing losses of the fully-supervised baseline and the fine-tuned model with early stopping for the downstream classification task.}
\label{fig:test_losses_performance}
\end{figure}

In terms of training efficiency, the self-supervised features can reach the fully-supervised performance quite quickly within 32 epochs of fine-tuning.
This is corroborated in Fig.~\ref{fig:test_losses_performance} that analyses the testing loss during fine-tuning against its fully-supervised counterpart.
This indicates that the self-supervised representation has indeed learnt good generalisable features.
We expect much greater data efficiency gains under more elaborate classification settings, such as a 1000-category classification task~\cite{henaff2020data}.
Note that unlike fully-supervised systems, the initial training of self-supervised features would amortise over many different downstream tasks.

\vspace{-0.20cm}
\subsection{Qualitative}
\begin{figure}[t]
  \centering
  \captionsetup{justification=centering}
  \subfloat[\footnotesize pre-trained VGG with\\ random head]{
    \includegraphics[width=0.215\textwidth]{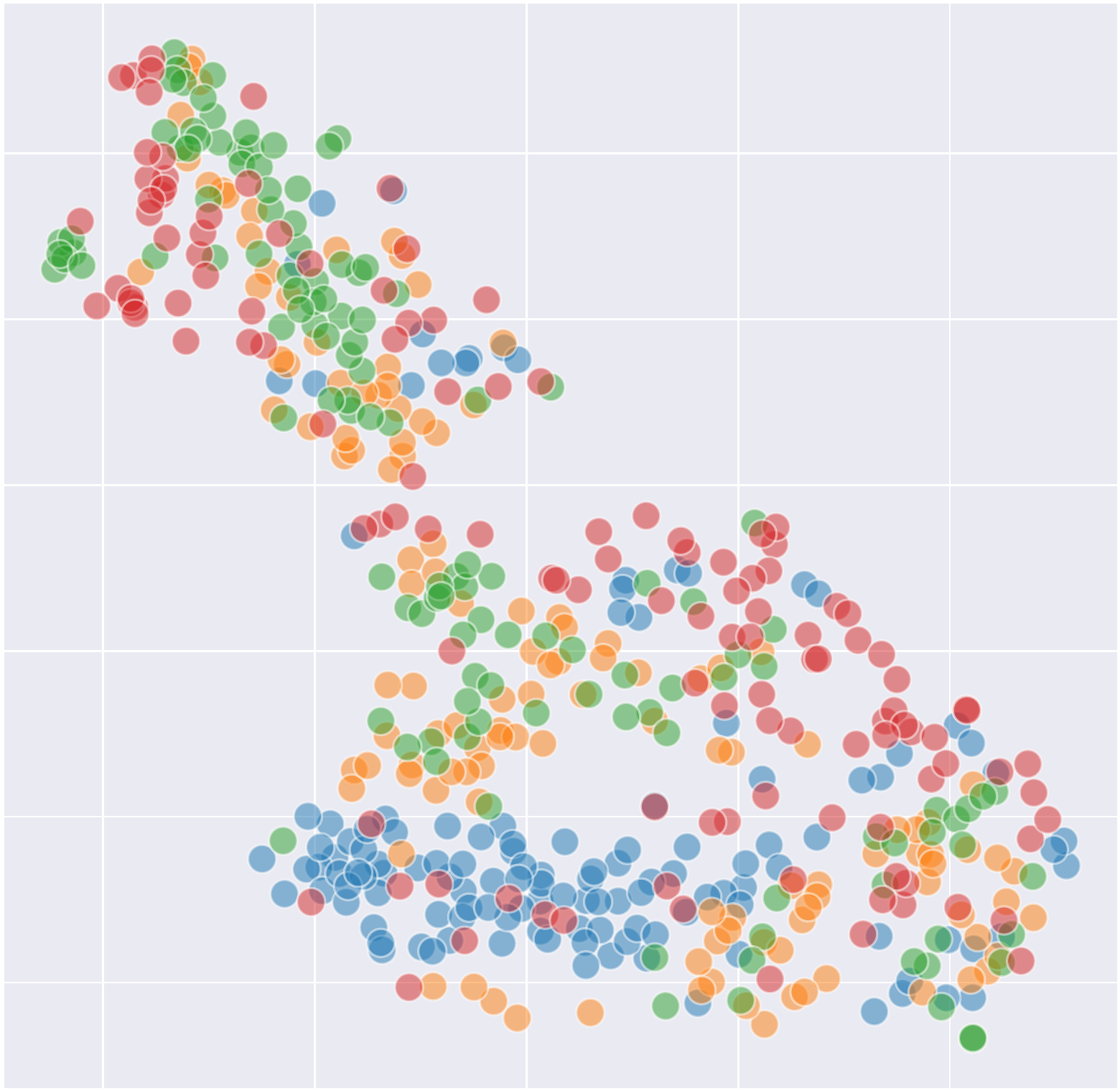}
    \label{fig:tsne_vgg}
  }
  \hspace{-0.35cm}
  \subfloat[\footnotesize fine-tuned contrastive\\ radio model]{
    \includegraphics[width=0.215\textwidth]{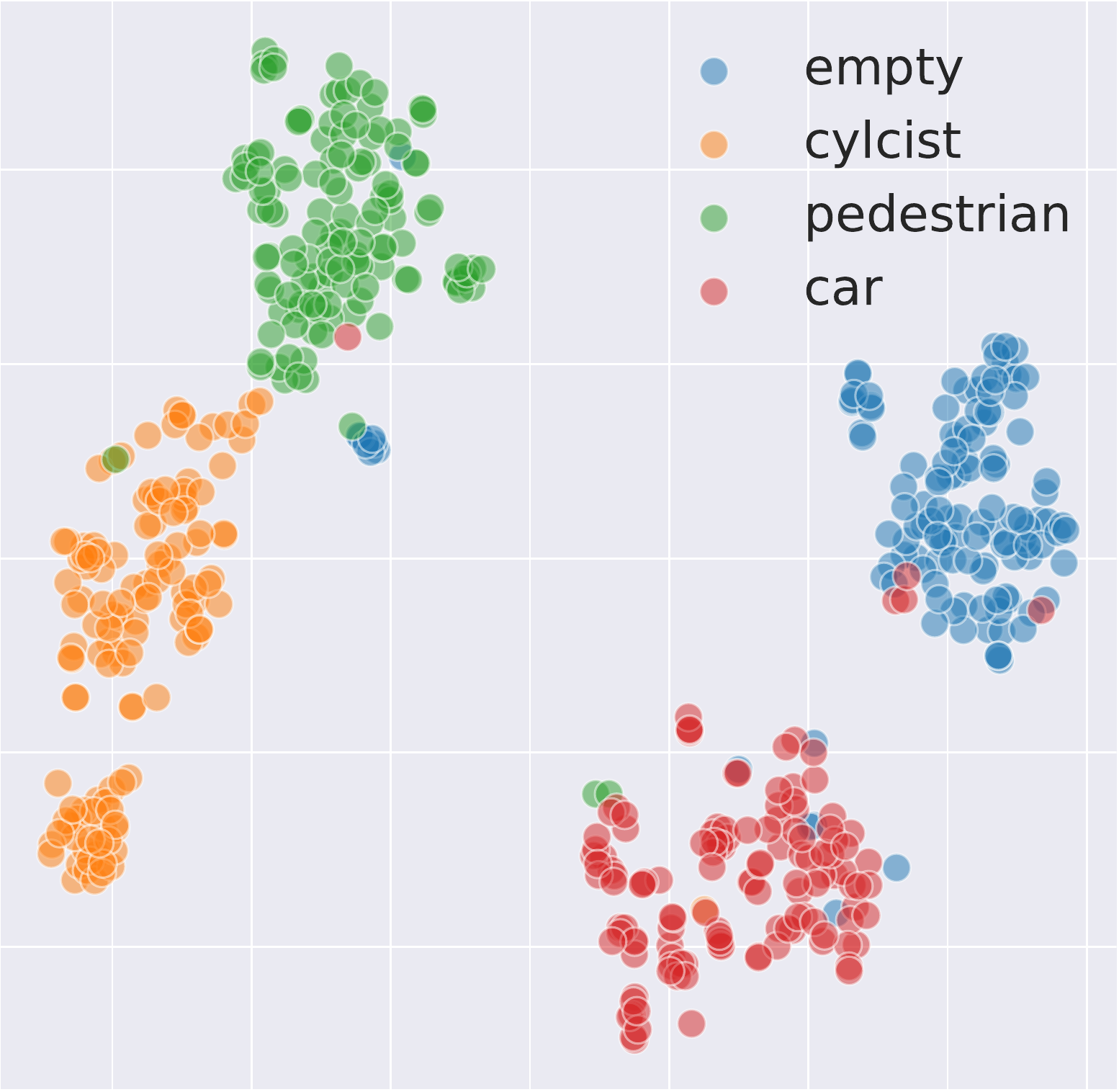}
    \label{fig:tsne_finetune}
  }
  \vspace{-0.15cm}
  \caption{t-SNE visualisation of learnt representations.} 
  \label{fig:qualitative_tsne}
\end{figure}

\noindent \textbf{Learnt representation.} As discussed in Sec.~\ref{sec:implementation_details}, we start the self-supervision of the radio branch from a pre-trained vision VGG network.
Could this, rather than contrastive training, be responsible for the very good performance on the downstream task?
After all, objects showing as blobs in radio heatmaps should in theory also benefit from many learnt vision filters, such as edge and contour detectors.
To investigate this, we use t-SNE~\cite{van2008visualizing} to visualise the 128-dimensional features of the penultimate layer for (a) a pre-trained VGG combined with randomly initialised classification head, and (b) our fine-tuned contrastive model.
Results are shown respectively in Fig.~\ref{fig:qualitative_tsne}.
Inspecting Fig.~\ref{fig:tsne_vgg}, no structure in the latent space can be discerned in relation to our 4-category downstream task.
In contrast, the latent space of our fine-tuned model exhibits clear clustering around the 4 categories of the downstream task as evident in Fig.~\ref{fig:tsne_finetune}.

\section{Related Work}

Self-supervised learning, where the model is trained on implicit labels present within the dataset, has gained popularity in the fields of NLP and CV recently. 
In this section we highlight some key related works.

\noindent\textbf{Computer Vision.} Self-supervised CV models trained with simple auxiliary tasks lead to surprisingly good vision representation learning.
Examples include solving puzzles~\cite{noroozi2016unsupervised}, image colorisation~\cite{zhang2016colorful}, and predicting image rotations~\cite{gidaris2018unsupervised}. 
For the latter for instance, self-supervised learning works by randomly rotating all the images in a dataset and asking the model to predict these rotations. 
Bleeding-edge self-supervision systems include SimCLR by Google~\cite{chen2020simple} or SwAV by Facebook AI Research~\cite{caron2020unsupervised}. 
SimCLR generates two augmentations for the same image and tries to maximise the similarities between their latent representations. 
SwAV builds on SimCLR and adds multi-crop augmentation, i.e., two copies of an image at two different resolutions. 
After training on auxiliary tasks, the learnt features are then fine-tuned for a given downstream task. 

\noindent\textbf{RF Sensing.} Self-supervised learning has recently been applied to radar systems research.
In~\cite{orr2021coherent}, the authors propose a super-resolution method called Radar signal Reconstruction using Self-supervision (R2-S2) which improves the angular resolution of a given radar array without increasing the number of physical channels. 
In ~\cite{gasperini2021r4dyn}, the authors propose R4Dyn which uses radars during training as a weak supervision signal, as well as an extra input to enhance the depth estimation robustness at inference time.

\section{Conclusion}

We asked if self-supervised learning can be used to train radio models using a radio-visual co-learning approach, and without relying on laboriously labelled data.
We have found that this is indeed not only possible, but is also superior to the performance of a fully-supervised baseline, especially when less amount of labelled data is used.
Our early findings are important for future 6G cellular systems where obtaining high-quality labelled data presents cost and scale challenges. 

For future work, we plan to build on our early results and investigate more elaborate self-supervised formulations. 
Chiefly, we would like to (i) incorporate temporal mutual information in self-supervision and (ii) expand our evaluation to a larger and richer radio-visual dataset whereupon we can implement more complex downstream tasks and expose yet more compelling benefits to self-supervised learning for radio systems.

\section*{Acknowledgment}
We would like to thank Murhaf Hossari for contributing to an early form of this research. We thank Howard Huang for his feedback on this manuscript.

\IEEEtriggeratref{15}
\bibliographystyle{IEEEtran}
\bibliography{references}

\newpage

\section{Connection between $\mathcal{L}_c$ and MI} \label{sec:Lc_MI_connection}

As per Eq.~\eqref{eq:unnormalised_density_ratio} in Sec.~\ref{sec:contrastive_loss}, the score function between $q$ and $k$ models an unnormalised density ratio.
The optimal value of this score function is when it captures the underlying density ratio between the unencoded signals $x$ and $y$ (see~\cite{oord2018representation})
\begin{align}
\exp{( q \cdot k /\tau )} \bigg|_{\text{optimal}} &= \frac{p(y|x)}{p(y)} \label{eq:unnormalised_density_ratio_optim}
\end{align}

Using Eq.~\eqref{eq:unnormalised_density_ratio_optim} in Eq.~\eqref{eq:contrastive_loss}, and denoting a set of $K$ negative $y$ signals by $Y^{-} = \{y_{i}^{-}\}_{i=0}^{K-1}$, we obtain
\begin{align}
\mathcal{L}_c^{\text{optimal}} &= - \underset{x,y}{\mathbb{E}}\log \left[ \frac{\frac{p(y|x)}{p(y)}}{\frac{p(y|x)}{p(y)} + \sum_{ y_i \in Y^{-}} \frac{p\left(y_i|x\right)}{p\left(y_i\right)}}\right] \notag \\
&= - \underset{x,y}{\mathbb{E}} \log \left[ \frac{1}{1 + \frac{p(y)}{p(y|x)} \sum\limits_{ y_i \in Y^{-}} \frac{p\left(y_i|x\right)}{p\left(y_i\right)}} \right] \notag \\
&= \underset{x,y}{\mathbb{E}} \log \left[ 1 + \frac{p(y)}{p(y|x)} \sum\limits_{ y_i \in Y^{-}} \frac{p\left(y_i|x\right)}{p\left(y_i\right)} \right] \notag \\
&\approx \underset{x,y}{\mathbb{E}} \log \left[ 1 + \frac{p(y)}{p(y|x)} (K-1) \underset{y_i}{\mathbb{E}} \frac{p\left(y_i|x\right)}{p\left(y_i\right)} \right] \notag \\
&= \underset{x,y}{\mathbb{E}} \log \left[ 1 + \frac{p(y)}{p(y|x)} (K-1) \right] \notag \\
&\geq  \underset{x,y}{\mathbb{E}} \log \left[ \frac{p(y)}{p(y|x)} K \right] \notag \\
&= - I(x;y) + \log(K) \notag
\label{eq:contrast_loss_optim_derivation}
\end{align}
\begin{flushright} 
\vspace{-0.66cm}
\qedsymbol{}
\end{flushright} 
%


\end{document}